\documentclass{IEEEtran}
\usepackage{cite}
\usepackage{amsmath,amssymb,amsfonts}
\usepackage{graphicx}
\usepackage[utf8]{inputenc}
\usepackage[export]{adjustbox}
\usepackage{wrapfig}
\usepackage{relsize}
\usepackage{textcomp,nicefrac}
\usepackage[switch, modulo,pagewise]{lineno}

\def\BibTeX{{\rm B\kern-.05em{\sc i\kern-.025em b}\kern-.08em
T\kern-.1667em\lower.7ex\hbox{E}\kern-.125emX}}
\newcommand{\neqcm}{$n \cdot \mathrm{cm}^{-2}\;$}
\begin{document}
\title{Radiation tolerance of online trigger system for COMET Phase-I}
\author{Sam Dekkers, Yu Nakazawa, Yuki Fujii, Hisataka Yoshida, Ting Sam Wong, Kazuki Ueno, Jordan Nash
\thanks{Manuscript received October 30, 2020; revised March 31, 2021;
accepted May 24, 2021.}
\thanks{This work was supported by JSPS KAKENHI Grant Numbers JP17H04841, JP18H05231 and 18J10962 and partially by the Australian Government through the Australian Research Council's Discovery Projects funding scheme (project DP200101562). The views expressed herein are those of the authors and are not necessarily those of the Australian Government or Australian Research Council.}
\thanks{S.Dekkers, is with the School of Physics and Astronomy, Monash University, Clayton, VIC 3168 Australia (e-mail: Sam.Dekkers1@monash.edu).}
\thanks{Y.Nakazawa, is with the Department of Physics, Osaka University, Osaka, Japan (e-mail: y-nakazawa@epp.phys.sci.osaka-u.ac.jp).}
\thanks{Y.Fujii, is with the School of Physics and Astronomy, Monash University, Clayton, VIC 3168 Australia (e-mail: Yuki.Fujii@monash.edu).}
\thanks{H.Yoshida, is with the Research Centre for Nuclear Physics, Osaka University, Osaka, Japan (e-mail: hisataka@rcnp.osaka-u.ac.jp).}
\thanks{T.S.Wong, was with the Department of Physics, Osaka University, Osaka, Japan (Graduated).}
\thanks{K.Ueno, is with Institute of Particle and Nuclear Studies, KEK, Tsukuba, Japan (e-mail: kazuueno@post.kek.jp).}
\thanks{J.Nash, is with the School of Physics and Astronomy, Monash University, Clayton, VIC 3168 Australia (e-mail: Jordan.Nash@monash.edu).}
}

\maketitle

\begin{abstract}
 The COMET experiment aims to search for the neutrinoless muon to electron transition process with new sensitivity levels.
 The online trigger system is an integral part of achieving the sensitivity levels required and will be subject to an expected neutron fluence of up to  $10^{12}$~\neqcm within regions inside the detector solenoid.
 Consequently a significant number of soft errors in the logic of the onboard field programmable gate arrays (FPGA) can occur, requiring error correction for single event upsets and firmware reprogramming schemes for unrecoverable soft errors.
 We studied the radiation tolerance of the COMET Phase-I front-end trigger system, called COTTRI, subject to neutron fluence on order $10^{12}$~\neqcm with multiple error correcting codes and automatic firmware reconfiguration.
 The regions measured were the configuration RAM, block RAM and also in a multi-gigabit transfer link using copper cables that will be used for communication between different trigger boards during Phase-I.
 The resulting cross sections observed suggest the most significant impact to the experiment will come from unrecoverable soft errors in configuration RAM, with dead time expected to be $(4.2 \pm 1.3)\%$.
 The effect of multi-bit errors in block RAM was found to be almost negligible in COMET Phase-I.
 In addition, multiple solutions have already been proposed in order to suppress these errors further.
 Soft errors observed in the multi-gigabit transfer links were measured to be of two orders of magnitude less impact compared to the unrecoverable errors in configuration RAM.
 We concluded that the COTTRI system meets the trigger requirement in COMET Phase-I. 
\end{abstract}


\section{Introduction}
\label{sec:introduction}
\IEEEPARstart{T}{he} COMET experiment (\textbf{CO}herent \textbf{M}uon to \textbf{E}lectron \textbf{T}ransition) is a physics experiment aiming to measure the charged lepton flavour violating process of $\mu+N\rightarrow e+N$, a process forbidden in the Standard Model of particle physics \cite{b1}. 
Any measurement of such an event would be clear evidence for new physics, with many models predicting rates of occurrence within the sensitivity levels that COMET is aiming for. Phase-I of the COMET experiment is aiming to reach sensitivity levels of approximately $3\times10^{-15}$, an improvement of 100 times on sensitivity limits reached by previous experiments measuring the same process \cite{b1}.
However to reach such sensitivity levels, every subsystem in the experiment needs to be optimised.
\newline
\par One critical subsystem is the field programmable gate array (FPGA) based COTTRI (\textbf{CO}ME\textbf{T} \textbf{TRI}gger) system being utilised in Phase-I of the COMET experiment.
This online trigger system will be responsible for choosing interesting physics events during data collection, and must do so within the harsh radiation conditions expected to be present within the detector region of the experimental facility. The neutron beam used for this experiment was chosen to best match the neutron conditions expected during Phase-I with the main aim of these tests being to measure the soft error cross sections to be observed specifically during Phase-I measurements.
\\
\\
In order to ensure sensitivity requirements will be met, one of the current COTTRI front-end boards was irradiated with neutron radiation levels up to those expected during Phase-I of the experiment. It is expected in some regions of the detector solenoid that a total neutron fluence of up to $10^{12}$~\neqcm, including a safety margin factor of 5 \cite{b2}, will be present over the data collection period. During irradiation, soft errors were measured in different sub-systems of the board; multi-gigabit data transfer links (MGT), configuration random access memory (CRAM) and block random access memory (BRAM).
The MGT behaviour is of particular interest as these tests have not been performed before in preparation for COMET Phase-I.
These results along with conclusions about their impact on data collection for COMET Phase-I will be discussed in this paper.
\section{Background}
\subsection{COMET and the COTTRI System}
The COMET Phase-I experiment is currently under construction at the Japan Proton Accelerator Research Complex (J-PARC) in Tokai, Japan and will utilise an 8 GeV pulsed proton beam directed at a graphite target in order to produce a high intensity pion beam.
Within the length of a transport solenoid, pions will decay into muons resulting in the highest intensity muon beam entering the detector region.
The CyDet (\textbf{Cy}lindrical \textbf{Det}ector) used in Phase-I will be centred around aluminium stopping targets in order to observe the capture and decay of muons into electrons within the presence of the aluminium nuclei.
Surrounding the aluminium stopping targets are a 1 T detector solenoid and CDC (\textbf{C}ylindrical \textbf{D}rift \textbf{C}hamber) detector with two CTH (\textbf{C}ylindrical \textbf{T}rigger \textbf{H}odoscope) detectors located at each end of the detector region.
The CDC detector will provide momentum measurements for signal electrons while the CTH detectors will provide important timing information.
The detector solenoid will direct high momentum charged particles to the CTH detectors at either end allowing for suppression of lower momentum backgrounds.
\newline
\par
\begin{figure}[h]
     \centering
     \includegraphics[scale=0.26]{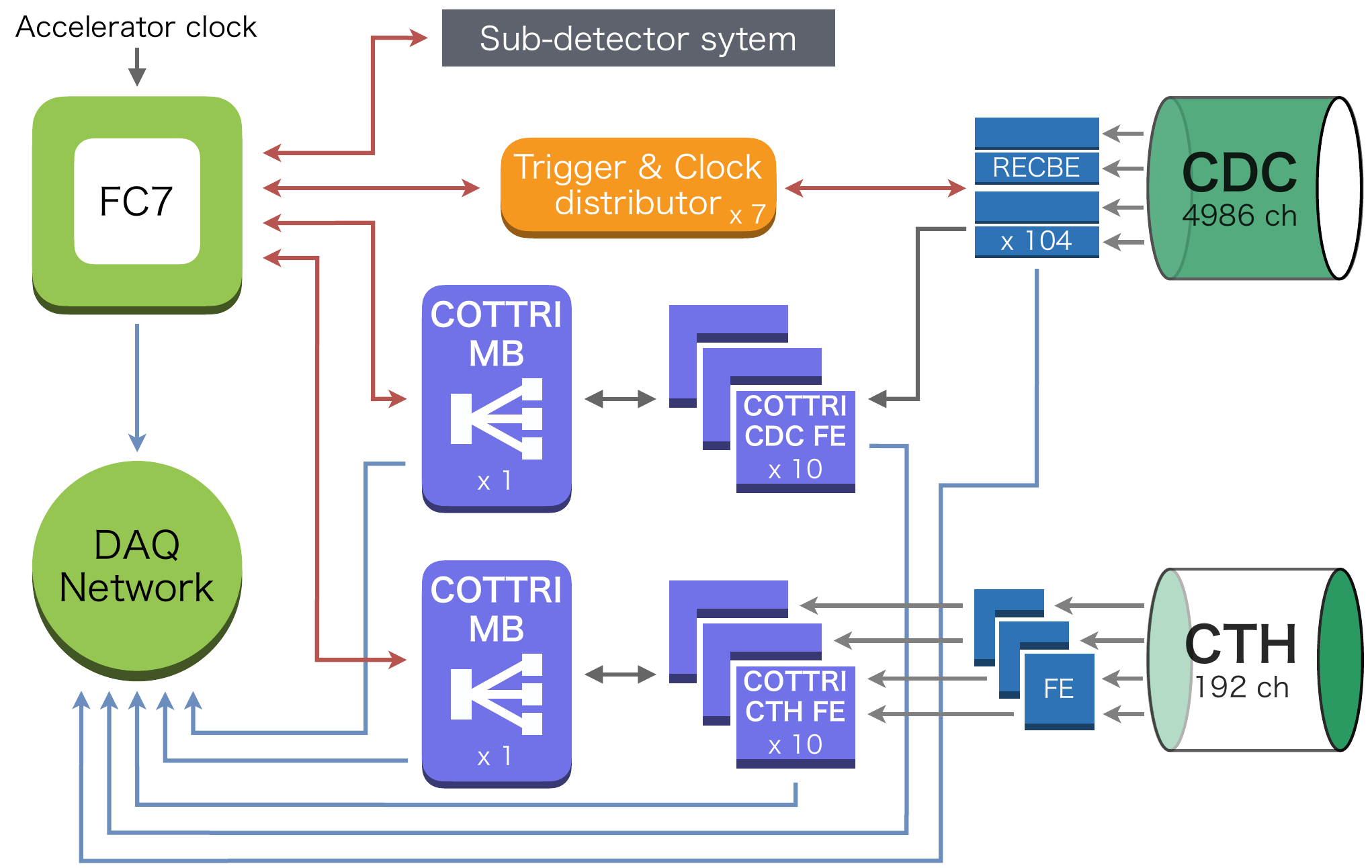}
     \caption{An overview of the COTTRI system for Phase-I of the COMET experiment \cite{b1}. The COTTRI-FE board used in this experiment is one of the 10 boards seen in this diagram connected to the CDC RECBE boards that will take information from the CDC for making trigger decisions.}
     \label{TriggerOverview}
\end{figure}

\begin{figure}[h]
     \centering
     \includegraphics[scale=0.60]{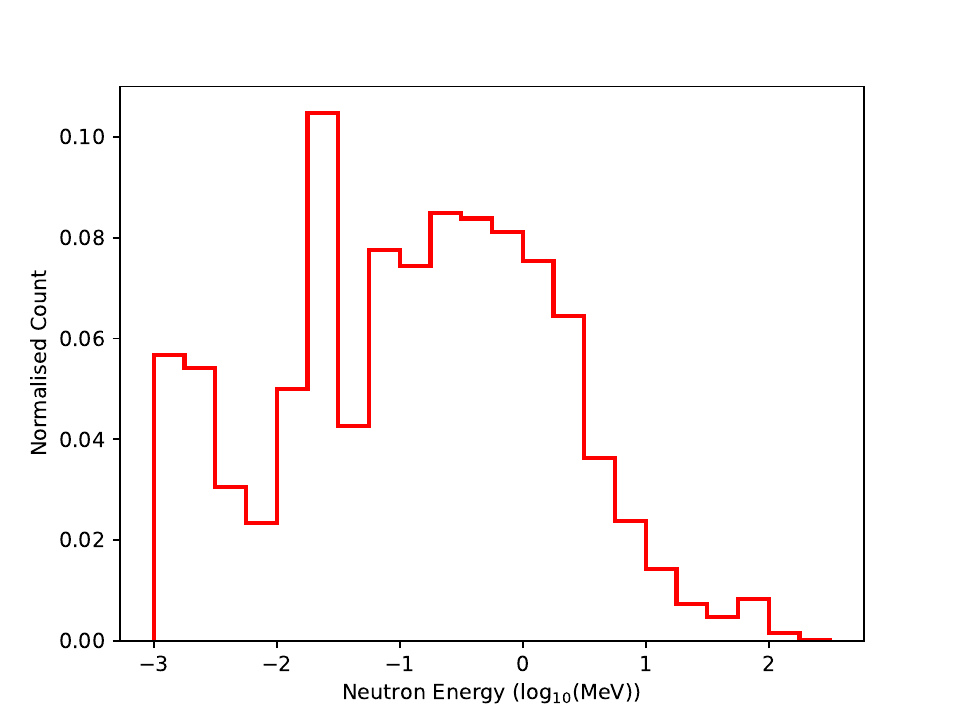}
     \caption{Simulated neutron radiation energy spectrum (produced with PHITS3.000 \cite{bPHITS}) expected at the entrance to the detector region for Phase-I of COMET. A majority of expected neutrons can be seen to be of energies below 10 MeV, leading to the choice of a lower energy beam facility for this study. However, the effects of the higher energy components of this energy spectrum are also considered in discussions on the expected SEFI rate in Section \ref{sec:Discussion}.}
     \label{NeutronSpectra}
\end{figure}

 The trigger system for this experiment is important in managing the high amount of event data that will be produced by the high intensity muon beam entering the detector region. One of the major components for the Phase-I trigger system is the COTTRI system which will take event data produced by the CDC and CTH detectors in order to make trigger decisions. An overview of the Phase-I trigger system can seen in Figure \ref{TriggerOverview}.
The CDC detector output will be connected to readout electronics boards (RECBE) (originally designed and developed for and by the Belle II experiment \cite{b3}) which will feed into 10 COTTRI Front-End (FE) boards located behind the RECBE boards inside the detector solenoid.
The trigger information from the 10 FE boards will then be fed into the COTTRI Merger-Board (MB) with all final trigger decisions being processed by a central FC7 based processor \cite{b4}.
The CTH detector will also produce a trigger system based on front end discriminator electronics fed into the COTTRI-CTH FE boards based on the time-to-digital converters and analogue-to-digital converters, and those information will be processed into the COTTRI-CTH MB.
All of these links between the CDC RECBE channels and the COTTRI FE and MB rely on fast MGT links so it was critical to test these as well as the FPGA electronics in terms of their radiation tolerance. 
\\
\\
Over the course of the 150 day Phase-I data collection, it is expected based on simulation results for the CyDet detector region that up to $10^{12}$~\neqcm neutron fluence and 1 kGy (Si) gamma radiation can be expected in some parts. Both of these simulation results incorporate safety factors of 5. Focusing on the neutron irradiation, an expected energy spectrum for the entrance to the detector region is shown in Figure \ref{NeutronSpectra}, which shows that a majority of the expected neutrons interacting with the COTTRI boards will be of energies lower than 10 MeV. Simulations were produced using PHITS3.000 \cite{bPHITS}. These neutron energies are indicative of the distribution expected in the entire detector region. This spectrum also shows 76\% of neutrons expected to be 1 MeV or less, and 96\% expected to be less than 10 MeV.  For this reason measurements in this study were conducted in a manner that better replicates the expected neutron energies in order to gain a better understanding of the average effects to be expected over the course of the Phase-I measurements. However, as the number of higher energy neutrons is not insignificant, an estimation of their contribution to the radiation impact is discussed in Section \ref{sec:Discussion}.
\subsection{Radiation effects in FPGA electronics}
As mentioned previously, the COTTRI trigger system must operate in high levels of neutron fluence over the course of data collection in Phase-I.
It is well studied that different types of radiation can induce a variety of soft errors in silicon based electronic materials, as well as the potential for hard errors which can cause permanent damage \cite{b5}.
The implications of these effects are of importance in many fields such as space electronics research in addition to high energy physics \cite{b6,b7}.
\newline
\par The basis of this study is specifically on determining the expected impact of neutron irradiation to the COTTRI FE boards for the 150 day data collection period of COMET Phase-I.
The main process via which high energy neutrons may impact the FPGA electronics used in the COTTRI FE boards is through interactions with the silicon materials to produce secondary charged particles.
These charged particles can induce electron-hole pairs that can change the logical state of transistor gates on the FPGA \cite{b5}, potentially leading to a change to a bit in memory.
The general term used for these induced soft errors is single event upsets (SEU).
If a single bit is changed in a memory location, referred to as a single bit upset (SBU), it generally can be fixed with some form of error-correction code (ECC).
However issues arise if this data is affected by multiple bit upsets (MBU) as an ECC can only detect that this has occurred but can not correct it.
Furthermore if a soft error occurs in the CRAM of an FPGA, it has the potential to inflict functionality changes to the FPGA operation.
These errors can be classified as single event functional interrupts (SEFI) and they require re-configuring the firmware in order to fix. 
\section{Experiment}
The experiment comprises of three main measurements; the cross sections observed in the CRAM, BRAM and MGT links due to neutron radiation. To take these measurements, one of the COTTRI FE boards to be used with the CDC was placed in front of a neutron source with a test pattern and multiple ECCs were utilised in the FPGA firmware to check for errors induced by the radiation. This section describes the physical setup and the firmware used as well as the DAQ and slow control system required.
\subsection{Measurement Setup}
\begin{figure}
     \centering
     \includegraphics[scale=0.10]{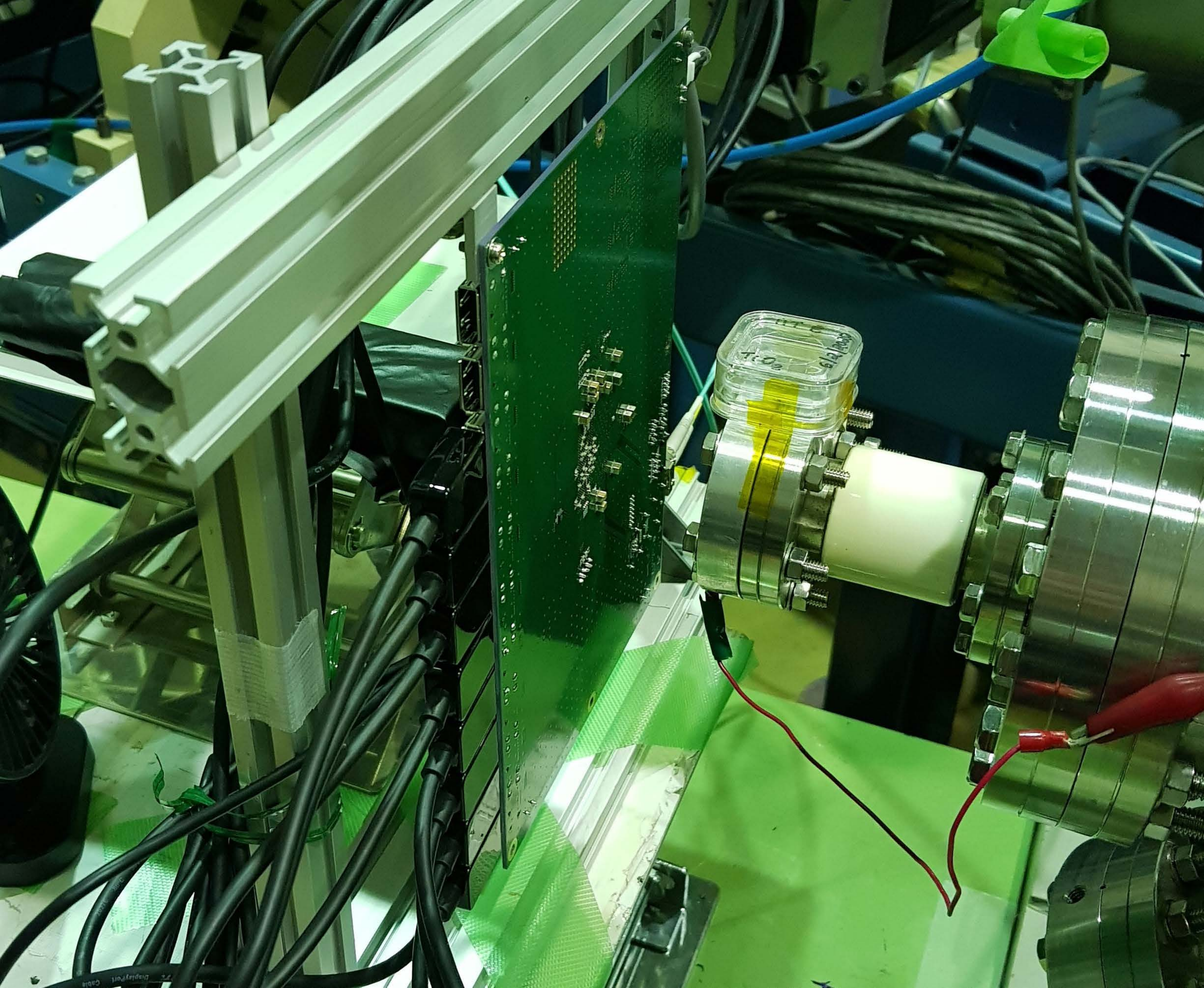}
     \caption{Image of COTTRI FE board located perpendicular to neutron source. The MGT link can be seen on the left side of the board in the image with the FPGA located on the reverse side of the PCB facing away from the neutron source on the right. The DAQ PC (not shown in this image) is located approximately 5 m away from this radiation zone.}
     \label{IMAGE}
 \end{figure}

\begin{figure}[t]
\includegraphics[scale=0.63]{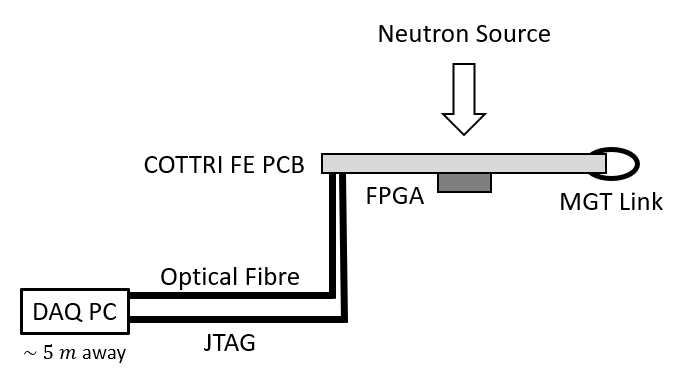} 
\caption{Diagram of setup (excluding power supply). The COTTRI FE board was placed perpendicular to the neutron source, centred on the FPGA that was facing away from the source. An optical fibre was utilised to send data to the DAQ PC every second while a JTAG connection was utilised for firmware download to the FPGA. The DAQ PC was situated outside of the radiation zone (approximately 5 m away).}
\label{PhysicalSetup}
\end{figure}

 Measurements were taken at Kobe University using their tandem electrostatic generator facilities to produce a 2 MeV peak neutron beam. The neutron beam is produced via a $^{9}\text{Be(d,n)} ^{10}\text{B}$ reaction by colliding a 3 MeV deuteron beam with a beryllium target. The neutron beam production setup and energy spectrum at this facility are equivalent to those discussed in \cite{bN}, with a higher energy component above 4 MeV of approximately 10\%. The characteristics of this neutron beam source are closer to the expected Phase-I neutron environment as shown in Figure \ref{NeutronSpectra} than other higher energy beam facilities, hence the choice of this facility.
 \\
 \\
 The COTTRI FE board used in testing was placed perpendicular to the neutron source at a measured distance as shown in Figure \ref{IMAGE}.
 Various distances were used over the duration of data collection; 5.15, 6.15, 8.15, 10.15 and 24.15 cm ($\pm$ 0.2 cm) which each include the COTTRI FE PCB thickness due to board orientation with respect to the neutron source. As previously studied in \cite{bPrevStudy}, the impact of the neutron radiation is asymmetric with respect to the board orientation. For this reason the orientation was chosen to provide the most conservative scenario when considering the placement of the COTTRI FE boards in the final COMET Phase-I design. The COTTRI FE board was connected to a DAQ PC located approximately 5 m away from the neutron source with both a JTAG cable for firmware configuration of the FPGA and an optical fibre for data transmission. Data transmission utilises an SiTCP \cite{b8} connection, a protocol for TCP/IP and UDP connections for devices such as FPGAs. An overview of this setup is shown in Figure \ref{PhysicalSetup}.
 \newline
 \par Over the data collection period, the COTTRI FE board was exposed to a total neutron fluence of $(1.0\pm 0.3)\times 10^{12}$~\neqcm.
 We calculated the neutron fluence levels through the use of conversion factors which were determined in previous studies and simulations \cite{bPrevStudy}. Using the distance and integrated beam current measurements in this study, the accumulative neutron fluence, $\mathlarger{\Phi_{\text{NF}}}$, is given by:
 \begin{equation}
     \mathlarger{\Phi_{\text{NF}}} = \sum_{\text{distance}} C_i\times I_i,
 \end{equation}
 where $I_i$ is the integrated beam current and $C_i$ is the conversion factor between integrated beam current and neutron fluence for a given perpendicular distance $d_i$ between the board and neutron source.
 The conversion factor $C_i$ is an averaged conversion factor coming from values in each position of an imposed grid over the FPGA surface area in order to account for angular dependencies. The value for the conversion factor at each angle $\theta_j$ is given by:
 \begin{equation}
     (C_i)_j = C_{10}\cdot(0.96-C_{\theta}\cdot\theta_j)\cdot\frac{10.0^2}{d_i^2}, \label{ANGLE}
 \end{equation}
 where $C_{10} = (4.9\pm 1.5)\;\text{MHz}\cdot$~\neqcm$\cdot \mu \text{A}$ \cite{bPrevStudy} is the conversion factor at 10 cm along the beam axis from the neutron source, $\theta_j$ is the angle made between the beam source and the centre of an imposed grid square, and $C_\theta$ is an empirical angular factor coming from the same previous measurements as $C_{10}$ \cite{bPrevStudy}. The increase in observed accumulative errors versus neutron fluence for all data sets (i.e. combining each distance measurement) was consistent when utilising the distance corrections outlined here.
\subsection{COTTRI CDC FE board}
\begin{figure}
    \centering
    \includegraphics[scale=0.45]{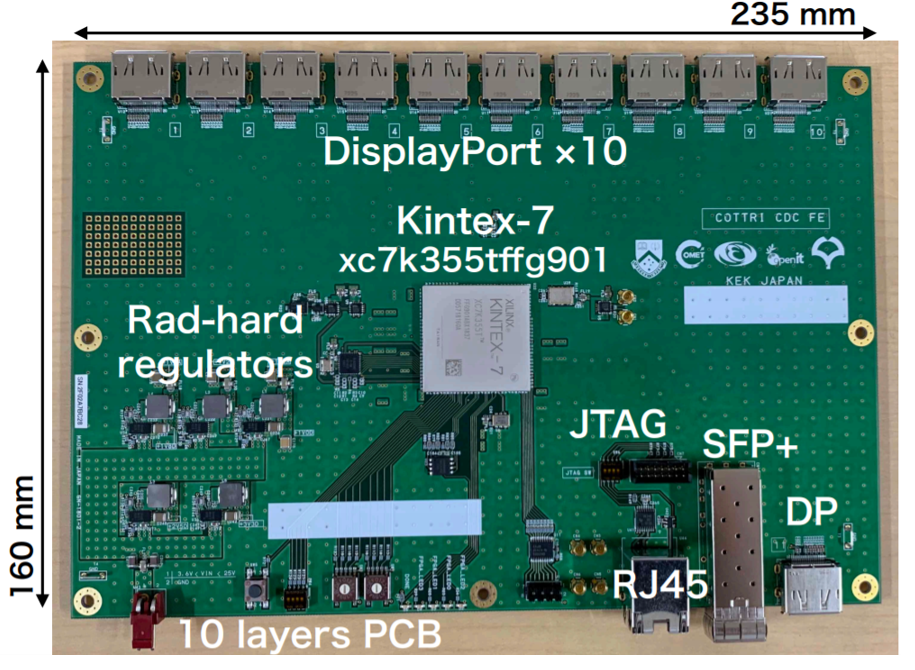}
    \caption{An image of one of the COTTRI CDC FE boards showing the main features including the Xilinx Kintex-7 series FPGA, display port connections used to construct the MGT link for measurements and the radiation hard voltage regulators important for operating the board in the radiation conditions expected during Phase-I data collection.}
    \label{BOARD}
\end{figure}

The COTTRI FE board used utilises a Xilinx Kintex-7 (XC7K355TFFG901) \cite{b10} and includes 10 display port connections, of which eight form the MGT link tested in the experiment by looping four 4.8 Gbps transfer cables between two ports each. The size of the FPGA chip itself is 3.1 $\text{cm}^2$, which is large enough to be reason for imposing a surface grid to include angular dependency effects on neutron fluence as outlined in (\ref{ANGLE}). An image of the board is shown in Figure \ref{BOARD}.
Other features in the COTTRI FE include radiation hard voltage regulators, a Joint Test Action Group (JTAG) connection and a small form-factor pluggable (STP+) connector for an optical fibre connection.

\subsection{Firmware}\label{sec:firmware}
\begin{figure}[t]
\includegraphics[scale=0.31]{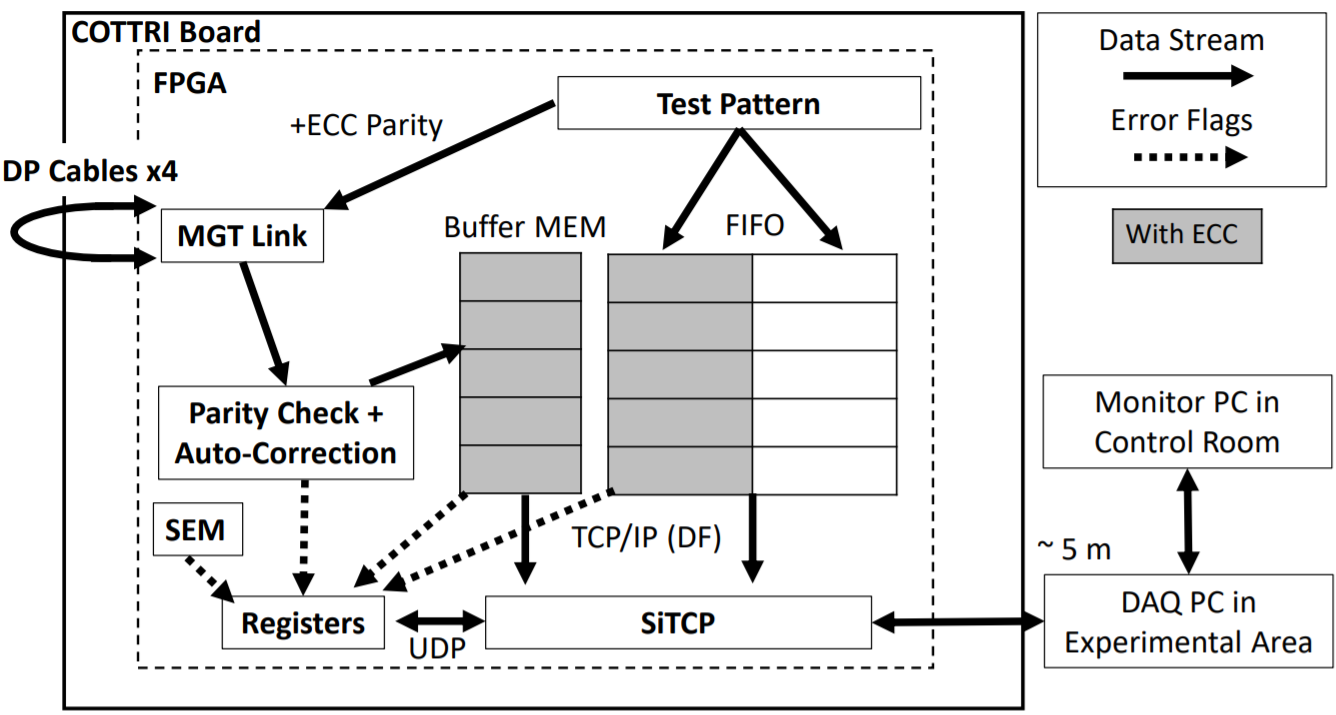} 
\caption{Data flow diagram for firmware on FPGA. A test pattern was sent to the MGT link that exits the COTTRI board via the four external DP loops that connect back to the board for error correction of SBUs (and record both SBU/MBU counts). Data is then sent to a ring buffer upon a 100 ns internal trigger. This buffered MGT link data is formatted with a data formatter (DF) into packets with test pattern data that was also separately sent to the FIFO blocks (half of which has the FIFO ECC applied) to be sent to the DAQ PC outside of the radiation zone utilising SiTCP.}
\label{FirmwareSetup}
\end{figure}

An overview of the firmware tested in this study is shown in Figure~\ref{FirmwareSetup}.
For the soft errors in CRAM, a Single Error Mitigation (SEM) IP Core \cite{b11} was utilised to automatically correct SEUs and detect SEFIs .
\newline
\par To estimate the soft errors in BRAM region, we prepared two fast-in-fast-out (FIFO) modules with and without built-in ECC module.
Identical test pattern data were injected into both FIFOs at the trigger and stored until the next trigger. The data pattern injected was a fixed 15 byte arbitrary non-zero test pattern.
Each FIFO has a length of $2^{16}$ and 64~bits per address, more than 25\%
of total memory size allowed in this FPGA chip to maximise the rate of soft errors in BRAM for the test purpose. The total FPGA resources utilised for this study are shown in Table \ref{RESOURCES}.
\newline
\par For the MGT links, we utilised the Aurora 8B/10B protocol \cite{b12} bypassing the memory buffer in order to ensure fast data transmission with a fixed latency less than one microsecond and a 4.8 Gbps data transfer rate. In COMET Phase-I the setup will be nearly identical for data transmission between two boards.
Before/after the data injection/extraction from the transceivers/receivers, ECC encoders/decoders are inserted to correct the SBUs and detect the MBU within one clock cycle.
The Aurora module provides other connection statuses such as soft error connection (SEC) and lost connection (LC), which are reported by the Aurora IP Core \cite{b12}. In both cases they automatically restart the MGT link within 1 s. 
For more validation and offline analysis, entire data from both BRAM and part of the data from the MGT link were sent to the DAQ PC through the SiTCP using TCP/IP protocol.
Along with these, all of the above error flags were fed into the registers and were updated so that they could be monitored through the UDP command via SiTCP, too. An outline of the firmware outlined in this section is shown in Figure \ref{FirmwareSetup}.
\begin{table}
\caption{FPGA Resource Utilisation}
\label{table}
\setlength{\tabcolsep}{3pt}
\centering
\begin{tabular}{|p{55pt}|p{55pt}|p{55pt}|p{55pt}|}
\hline
Resource&
Utilisation&
Available&
Utilisation \%\\
\hline
LUT&
19180&
222600&
8.62 \%\\
\hline
LUTRAM&
2311&
81400&
2.84 \%\\
\hline
FF&
29655&
445200&
6.66 \%\\
\hline
BRAM&
190.50&
715&
26.64 \%\\
\hline
IO&
43&
300&
14.33 \%\\
\hline
GT&
21&
24&
87.50 \%\\
\hline
BUFG&
18&
32&
56.25 \%\\
\hline
MMCM&
1&
6&
16.67 \%\\
\hline
PLL&
2&
6&
33.33 \%\\
\hline
\end{tabular}
\label{RESOURCES}
\end{table}

\subsection{DAQ and Slow control}\label{sec:daq}
The last component to the setup for this study was the DAQ and slow control that was implemented by a PC outside of the radiation zone as shown in Figure \ref{PhysicalSetup}.
The DAQ PC received error status and counter values via UDP connection to monitor the behaviour every 1 second, allowing us to check any correlations between different parameters by offline analysis.
Temperature, voltage levels and test pattern data were also monitored by the DAQ/slow control system.
\newline
\par A SEFI flag is given by the SEM module implemented inside the firmware.
In addition, we defined a few other SEFI behaviours that could not be detected by using the SEM module as follows:
\begin{enumerate}
    \item Connection loss - if either UDP or TCP/IP connection was lost, the event was counted as a SEFI.
    \item Erratic counter behaviour - if the counter value changed too rapidly this was also treated as a SEFI.
\end{enumerate}
Any of these behaviours were flagged as a SEFI and triggered a firmware reconfiguration immediately through a JTAG cable with a speed of 30~MBbs, that took approximately 20 seconds including the DAQ and monitor restarting.
The DAQ PC also received data sent via SiTCP connection every 1 s for offline analysis as mentioned in Sec~\ref{sec:firmware}.
\\
\\
No remote power cycling was implemented in this measurement setup, and it was required twice that manual power cycling be performed. This naturally will be a feature to implement in  future into the COTTRI system.

\section{Analysis and Results}\label{sec:Results}
This section outlines the resulting cross sections observed in each section of the COTTRI FE board. One key observation critical to performance was that no permanent errors or damage were observed in any region of the COTTRI FE board throughout this study.
\subsection{CRAM and BRAM}
The accumulative errors in the CRAM were counted based on flags returned from the SEM module while the BRAM errors were counted based on those returned from the FIFO ECC. It is important to note that the FIFO ECC only applied to half of the BRAM as its operation was validated based on the data sent offline via the SiTCP connection. Connection losses between the DAQ PC and COTTRI board were also counted as SEFI. Erratic behaviour in the SEU or SEFI counters were also counted as SEFI. Errors in CRAM and BRAM were accumulated and the resulting cross sections are shown in Table \ref{tabCRAMBRAM}. 
\begin{table}[h]
\caption{Cross Sections in CRAM and BRAM}
\label{table}
\setlength{\tabcolsep}{3pt}
\centering
\begin{tabular}{|p{53pt}|p{35pt}|p{68pt}|p{76pt}|}
\hline
Memory Region&
Error Type&
Cross Section (cm$^{2}$)&
Cross Section (cm$^{2}$/bit)\\
\hline
CRAM&SEU&
$(5.0\pm 1.6)\times 10^{-8}$&
$(6.4\pm 2.1)\times 10^{-16}$\\
&SEFI&
$(7.9\pm 2.6)\times 10^{-10}$&
$(1.0\pm 0.3)\times 10^{-17}$\\
\hline
BRAM&SBU&
$(1.8\pm 0.6)\times 10^{-9}$&
$(7.0\pm 2.3)\times 10^{-17}$\\
&MBU&
$(5.3\pm 1.7)\times 10^{-10}$&
$(2.1\pm 0.7)\times 10^{-17}$\\
\hline
\end{tabular}
\label{tabCRAMBRAM}
\end{table}
\newline
\par
\begin{figure}[h]
    \centering
    \includegraphics[scale=0.57]{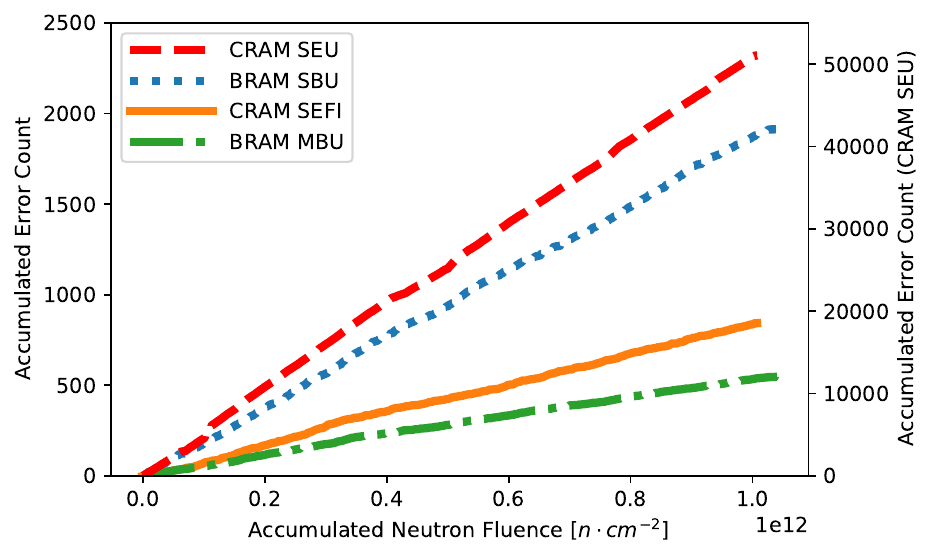}
    \caption{Accumulated error count plotted against accumulated neutron fluence for the CRAM and BRAM error counts. The scale for CRAM SEU error counts is shown on the right vertical axis.}
    \label{LINEAR}
\end{figure}

 The error counters for both CRAM and BRAM were observed to increase linearly with neutron fluence levels as shown in Figure \ref{LINEAR}. This figure visualises the linear trend of accumulated errors with accumulated neutron fluence. This is important as it shows that the systematic uncertainty during the measurements was relatively small even with instability in the neutron beam and changes to the position of the COTTRI FE board with respect to the neutron source.
\newline
\par The cross sections shown in Table \ref{tabCRAMBRAM} are consistent with the previous measurements performed with a previous COTTRI FE board that utilised a different Kintex-7 FPGA \cite{bPrevStudy} when differences between the FPGA used are considered i.e. logic cells and surface area. These measurements are also consistent with recent literature such as the energy dependent cross sections measured in \cite{bCSE}.
\subsection{MGT Link}\label{sec:MGT}
The MGT link parity check and auto-correction ECC was used to return counts for SBU and MBU occurring in each display port as well as their SEC and LC status. The behaviour of the MGT link indicated by each of these error counts was accounted for when considering their impact.
For example when a SEFI occurred, the first ten seconds upon reprogramming the FPGA firmware were not included as this was conservatively considered dead time due to the necessary time required for the DAQ to be operational again. This also includes a safety buffer of a few seconds.
Observations of behaviour where the MBU count increased more than half of the SBU count were also considered to be anomalous, contributing to the overall dead time. The origin of the anomalous SBU/MBU counter increases is unknown and may affect data transferred over these MGT links. However, as these anomalies self recover within 1 s in the Aurora 8B/10B module, it is thought that the reason for their occurrence is different from other erratic counter behaviour observed in the SEM module that require a firmware reset to recover from. This may suggest that the problem is either in the comma alignment or is purely a timing based problem such that the serial/parallel data encoding/decoding interface loses perfect timing alignment for a short period.
\newline
\par Another consideration when analysing the error counters were undetected SEFI type behaviour inducing erratic changes in the MGT link counters, different to the previously mentioned anomalous behaviour as they remained until a SEFI induced firmware reset. These were observed ten times over total data collection. In Section \ref{sec:Discussion} when considering the effects of SEFI on dead time, these errors will be included as they are a newly observed error that should be recognised as a SEFI by the DAQ, just as is observed in the CRAM. Taking into account the aforementioned behaviours, the cross sections observed in the MGT link are $(7.1\pm 2.3)\times 10^{-10}$ and $(2.1\pm 0.9)\times 10^{-12}~$cm$^{2}$ for SBU and MBU errors respectively.
\newline
\par Unlike the cross sections observed in the CRAM and BRAM, the errors observed in the MGT link occurred less frequently and not at a constant rate over time - it was observed that the increases in the SBU and MBU counters were within 1 s with this time being limited by the data acquisition rate.
The majority of observed errors were SBU errors which were resolved by the ECC. 
Additionally, any MBU increase or previously mentioned anomalous could be considered as 1 s of deadtime (also limited by the frequency of the DAQ system logging counters) when considering how the MGT link might be impacted during Phase-I data collection. 
\newline
\par The SEC and LC errors were also recorded in terms of the time impact they will have on data collection since any occurrence of such must be considered dead time during Phase-I data collection.
The time period during measurements that saw an increase in these three counters (with the MBU deadtime here including the anomalous behaviour previously mentioned) is shown below in Table. \ref{tabMGT2}.
\begin{table}[h]
\caption{MBU, SEC and LC Deadtime in MGT Link}
\label{table}
\setlength{\tabcolsep}{3pt}
\centering
\begin{tabular}{|p{95pt}|p{75pt}|}
\hline
Error Type&
Deadtime (s)\\
\hline
SEC&
$(4.4\pm0.7)\times 10$\\
LC&
$(4.3\pm0.7)\times 10$\\
MBU&
$(1.8\pm0.4)\times 10$\\
\hline
Total (excluding overlaps)&
$(6.5\pm0.8)\times 10$\\
\hline
\end{tabular}
\label{tabMGT2}
\end{table}
\newline
\par As with the SBU and MBU counters, time periods of increase in these counters in the 10 s immediately after a CRAM SEFI are not included as they are considered within the SEFI dead time required for a firmware reset. So as shown in Table \ref{tabMGT2}, the deadtime due to the SEC and LC counters along with the deadtime created by MBU counter increases (regardless of whether the SBU counter has increased or not) tallies to $(6.5\pm0.8)\times 10$ s over the entire MGT link data set. It should be noted that this total takes into account overlaps in counter increases - for example, if the SEC and LC counters both increase during the same DAQ period, this only contributes 1 s of deadtime to the total).
\section{Discussion}\label{sec:Discussion}
With these measured cross sections it is important to determine the impact they will have when it comes to Phase-I during data collection. The MGT link results comparatively show less impact than other regions with the MBU cross section being two orders of magnitude lower than the CRAM SEFI cross section. It was also observed that the parity check module corrected most SBU observed in the MGT link with SEC and LC also creating minimal dead time compared to the time period of data collection. With a total deadtime due to MBU, SEC and LC counter increases of $(6.5\pm0.8)\times 10$ s, the deadtime impact for 10 FE boards based on this value can be expected to be $(6.5\pm0.8)\times 10^2$ s during Phase-I data collection. Compared to the 150 days of data collection, this is of relatively small impact ($5\times 10^{-3}\%$). For the MBU occurrences tagging could be utilised in future for the time periods in which the count increases. The most significant impact observed in the MGT link however was the previously mentioned ten observed erratic counter behaviour time periods (not to be confused with the anomalous behaviour previously described and included in the  $(6.5\pm0.8)\times 10^2$ s of MGT link deadtime) which were corrected upon a CRAM SEFI induced firmware reset. In future this behaviour should be monitored and treated the same by the DAQ as a SEFI just as is done with similar CRAM behaviour.
\newline
\par Based on the observed cross sections in each FPGA region, the CRAM SEFIs will pose the most significant issue for data collection in terms of deadtime.
To estimate the impact on Phase-I data collection, the dead time based on expected SEFI numbers over 150 days is determined.
For 10 FE boards over the 150 days of data collection and exposed to the maximum expected $10^{12}$~\neqcm  neutron fluence, the measured CRAM SEFI cross section of $(7.9\pm 2.6)\times 10^{-10}$ cm$^{2}$ will correspond to $(7.9\pm 2.6)\times10^3$ total SEFI. The uncertainty in this result is dominated by the systematical one on the absolute neutron fluence.
The SEFI-like behaviour in the MGT can also be included and added to this total, corresponding to a further $(1.0\pm0.3)\times10^2$ total SEFI. Based on these total SEFI errors we can expect a SEFI rate of $(2.2\pm 0.7)\;\text{hr}^{-1}$ during the 150 day Phase-I data collection. The dead time observed for the one FE board in this study was approximately 30 s which includes the 10 s time period for allowing the DAQ to reset mentioned in Sec \ref{sec:MGT}. However a more conservative estimate is used as it is possible during the Phase-I experiments that the DAQ PC will not be located close to the COTTRI system along with other issues that may come with the final design. The conservative estimate utilised for the expected dead time per SEFI is 60 s.  With this SEFI dead time and the $(2.2\pm 0.7)\;\text{hr}^{-1}$ SEFI rate, the total expected dead time over the 150 days using this measurement is $(3.7\pm 1.2)\%$.
\\
\\
However this value is a measurement based on the neutron beam conditions provided by the Kobe University facility. To make a better estimate on the expected rate of SEFI during COMET Phase-I, $R_{\,\text{SEFI}}^{\,\text{COMET}}$, the expected neutron energy spectrum from Figure \ref{NeutronSpectra} and the neutron energy spectrum produced by a 2.8 MeV deuteron beam on a beryllium target from \cite{bN} can be combined with literature measurements on SEU cross-sections at higher energies such as in \cite{bCSE}. An estimate on the expected rate $R_{\,\text{SEFI}}^{\,\text{COMET}}$  based on the difference between the two neutron energy spectra is provided by:
\begin{equation} \label{CS}
    R_{\,\text{SEFI}}^{\,\text{COMET}} = \frac{\int F_{\text{COMET}}(\text{E})W(\text{E})\text{dE}}{\int F_{\text{KOBE}}(\text{E})W(\text{E})\text{dE}}R_{\,\text{SEFI}}^{\,\text{KOBE}},
\end{equation}
where $F_{\text{COMET}}$ and $F_{\text{KOBE}}$ are the neutron energy spectra for each facility environment, $W(E)$ is the SEU cross section dependence on neutron energy based on measurements made in \cite{bCSE} and $R_{\,\text{SEFI}}^{\,\text{KOBE}}$ is the SEFI rate measurement made in this study. The resulting estimated SEFI rate from (\ref{CS}) is then $R_{\,\text{SEFI}}^{\,\text{COMET}} = (2.5 \pm 0.8)~\text{hr}^{-1}$ which equivalently leads to a dead time over the 150 day Phase-I of $(4.2\pm 1.3)\%$.
\\
\par For the purposes of this study and Phase-I, a trigger efficiency of 90$\%$ is a target for the sensitivity limits of the experiment. To evaluate the trigger efficiency in the COTTRI system, we can consider an overall trigger efficiency: 
\begin{equation}
    \epsilon_{\text{Total}} \approx \epsilon_{\text{Trig}}\times \epsilon_{\text{Op}},
\end{equation}
where $\epsilon_{\text{Trig}}$ is the online trigger efficiency and $\epsilon_{\text{Op}}$ is the total operation efficiency, taking into account the impact of the total expected dead time established in these measurements. The estimated online trigger efficiency is expected to be 96$\%$ \cite{b13}, so when taking the conservative upper limit of the SEFI dead time estimate made using (\ref{CS}) ($5.5\%$ which means $\epsilon_{\text{Op}} = 0.95$) with this trigger efficiency, the target trigger efficiency is fulfilled ($\epsilon_{\text{Total}} \approx 0.96\times 0.95 = 0.91$). 
If Poisson statistics are assumed and the upper bound on a 90\% confidence interval for dead time is used, this estimated trigger efficiency becomes $\epsilon_{\text{Total}} \approx 0.90$.
Even in the worst case scenario, the COTTRI system will satisfy the target trigger efficiency.
\newline
\par When it comes to the BRAM MBU cross section, it can be seen that it is of the same order of magnitude when compared to the CRAM SEFI cross section. One of the impacts these errors can have on Phase-I data collection is causing an inconsistency between the offline trigger data and that of actual data processed online to make the trigger decisions. To determine the expected impact, we can consider the size of each trigger, $S_T$, to be given by:
\begin{equation}
    S_T = N_{\text{RC}} \times N_{\text{ADC}} \times N_{\text{CLK}}, \label{size}
\end{equation}
where $N_{\text{RC}}$ is the number of RECBE channels per COTTRI CDC FE board, $N_{\text{ADC}}$ is the number of bits for each channel that the RECBE generates based on the CDC detector signal and $N_{\text{CLK}}$ is the number of clock cycles per trigger. In the Phase-I COTTRI design, the number of RECBE channels will be 480, the number of bits generated by each channel is 2 and the number of clock cycles per trigger is 10 \cite{b13}. Using (\ref{size}) we can calculate a trigger size of 9600 bits. The expected number of errors in the Phase-I experiment, $N^{\text{Exp}}_{\text{Errors}}$, based on the measured BRAM MBU cross section can then be determined by:
\begin{equation}
    N^{\text{Exp}}_{\text{Errors}} = R_{\text{MBU}} \times \mathlarger{\Phi_{\text{NF}}} \times\frac{S_T}{S_{\text{BRAM}}},
\end{equation}
where $R_{\text{MBU}}$ is the measured BRAM MBU cross section and $S_{\text{BRAM}}$ is the size of the BRAM. Given the measured MBU cross section when considering 10 COTTRI FE boards of $(5.3\pm 1.7)\times 10^{-9}\;$ cm$^{2}$, the accumulated neutron fluence of $(1.0\pm 0.3)\times 10^{12}$~\neqcm and the size of the BRAM in this study to which the FIFO ECC was applied of $64\times32767$ bits, this leads to an expected $(2.4\pm0.8)\times 10$ trigger errors over the entire Phase-I data collection. Considering a trigger rate of 13 kHz over the 150 days of measurement \cite{b13}, the impact expected of the BRAM MBU cross section is negligibly small, even without considering any expected higher energy neutron components.
\newline
\par When discussing the impact due to the SEFI rate, it is important to note that this is a worst case scenario - the SEFI dead time estimation incorporates the safety factor of five from the neutron radiation simulations, a conservative estimate of 60 s dead time per SEFI is used and also the COTTRI FE boards are assumed to be located in the most neutron radiation affected area of the COMET Phase-I detector region. However given that in the final detector design the neutron fluence will most likely be less than in this study (one design choice to reduce neutron effects already is orientating the COTTRI FE boards to face the FPGA in direction of highest neutron source, which was shown to be signifcant in \cite{bPrevStudy}) and that the dead time per SEFI is a conservative upper limit, it is expected that SEFI impact will be lower than determined in this study. In addition there are some simple improvements that can be implemented in order to suppress some of these soft errors even further. Implementing smaller block sizes within the BRAM has the potential to decrease BRAM MBU numbers by at least an order of magnitude, while triple modular redundancy is another potential solution although a resource intensive one.
For the observed MGT link behaviour, labels can be used with the anomalies described in Section \ref{sec:Results} so that the corresponding datasets taken during such occurrences can be discarded offline. Firmware can also be updated to treat the erratic SEFI-like counter behaviour in the MGT links in the same manner as other SEFI occurrences. As for the more impactful CRAM SEFIs, further improvements to the DAQ firmware reprogramming can also be implemented such as partial reprogramming to only correct affected regions and further reduce dead time per SEFI. Another consideration is that the implications outlined in this discussion assume that the dead time will affect all of the COTTRI FE boards. Instead it is possible for one board to be masked until it recovers while the other boards continue with data collection so that entire events may not be affected. 

\section{Conclusions}
The COMET experiment is searching for neutrinoless muon to electron conversion at new sensitivity levels, and requires the online COTTRI trigger system to be tolerant to high neutron radiation levels. 
This was studied by exposing the board to an accumulative neutron fluence of $(1.0\pm 0.3)\times 10^{12}$~\neqcm which is equivalent to the maximum expected fluence within the detector region during Phase-I including the safety factor of 5.
We estimated a total deadtime to be $(4.2\pm 1.3)\%$ during Phase-I due to all SEFI-type behaviour caused by neutron radiation effects. As most of them can be detected by an SEM module, partial reconfiguration and masking of an individual COTTRI FE board would be able to further suppress this deadtime.
Although cross sections within the BRAM and MGT links were insignificant compared to the SEFI one, several simple solutions have been already proposed as written in the previous section.
We will implement a data tagging function to recognise short time anomalous behaviour in MGT link to further improve the trigger efficiency before the start of COMET Phase-I. It will be important in future work to investigate the MGT link anomalous behaviour observed in this study further given that these MGT links form a critical component of the COTTRI system. However as a current status, by using the multiple ECC codes together with the firmware reconfiguration schemes, the COTTRI system will meet the requirements for COMET Phase-I.

\section{Acknowledgements}
The authors are grateful to Prof. Y. Furuyama, Dr. A. Taniike, Dr. Y. Mori and Mr. H. Kageyama (Kobe University) for the operation of the tandem electrostatic accelerator. 
The authors also thank Monash University for the financial support to accomplish this study.

\end{document}